\documentclass[5p,10pt,preprint]{elsarticle}
\usepackage{graphicx}
\usepackage{amssymb}
\usepackage{amsthm}
\usepackage{amsmath}
\usepackage{hyperref}
\usepackage{natbib}

\usepackage{xcolor}
\usepackage{amsmath}

\journal{Nuclear Materials and Energy}

\begin{document}

\begin{frontmatter}
\title{On the classification and quantification of crystal defects after energetic 
bombardment by machine learned molecular dynamics simulations}

\author[ipp]{F. J. Dom\'inguez-Guti\'errez}
\ead{javier.dominguez@ipp.mpg.de}
\cortext[cor1]{Corresponding author}
\author[uh]{J. Byggm\"astar}
\author[uh,hip]{K. Nordlund}
\author[uh,hip]{F. Djurabekova}
\author[ipp]{U. von Toussaint}

\address[ipp]{Max-Planck Institute for Plasma Physics, Boltzmannstrasse 2, 85748
 Garching, Germany.}
 \address[uh]{Department of Physics, University of Helsinki, Helsinki, PO Box 43, FIN-00014, Finland.}
 \address[hip]{Helsinki Institute of Physics, Helsinki, Finland.}

\begin{abstract}
The analysis of the damage on plasma facing materials (PFM), 
due to its direct interaction with the plasma environment, 
is needed to build the next generation of nuclear machines, 
where tungsten has been proposed as a candidate.
In this work, we perform molecular dynamics (MD) simulations 
using a machine learned inter-atomic potential, 
based on the Gaussian Approximation Potential framework, to 
model better neutron bombardment mechanisms in pristine W 
lattices.
The MD potential is trained to reproduce realistic 
short-range dynamics, the liquid phase, and the material 
recrystallization, which are important for collision cascades.
The formation of point defects is quantified and classified by a 
descriptor vector (DV) based method, which is independent of the 
sample temperature and its constituents, requiring only modest 
computational resources. 
The locations of vacancies are calculated by the k-d-tree 
algorithm.
The analysis of the damage in the W samples is compared to results 
obtained by EAM Finnis-Sinclair and Tersoff-ZBL potentials, at a 
sample temperature of 300 K and a primary knock-on atom (PKA) energy 
range of 0.5-10 keV, where a good agreement with the reported number 
of Frenkel pair is observed.
Our results provide information about the advantages and limits 
of the machine learned MD simulations with respect to the standard ones.
The formation of dumbbell and crowdion defects as a function of 
PKA is discussed.

\end{abstract}

\end{frontmatter}


\section{Introduction}
\label{intro}

The materials of the first wall of a fusion machine
have to deal with several challenging conditions.
For instance, a plasma facing material (PFM) is 
exposed to a hostile environment due to the plasma interaction,
high temperatures, and energetic neutron irradiation, to 
mention a few \cite{ZINKLE20112, EHRLICH200079}.
Tungsten has been used as a PFM due to its physical and 
chemical properties like low erosion rates, small tritium 
retention, and high melting point \cite{Marian_2017}.

In order to design the next generation of fusion machines, the 
analysis of different types of material defects in PFM is necessary
to better understand the effects of plasma irradiation on several 
physical and chemical properties of the materials. 
When an atom of the sample materials receives 
a higher kinetic energy
than the threshold displacement energy \cite{Luc75,Nor05c} 
it can produce permanent point or extended defects \cite{Federici_2001,wirth_hu_kohnert_xu_2015}.
For example, it has been observed that at high
initial energies of the primary knock-on atom (PKA),
defect clusters can be formed directly in crystalline 
materials \cite{Marian_2017,Sand_2013},
whereas simple point defects like self-interstitial-atoms (SIA)
are commonly formed at low impact energies.
In body-center-cubic (bcc) W lattice samples, SIA's are commonly
observed in an arrangement known as dumbbells and crowdions \cite{PhysRevB.73.020101}.


Molecular dynamics (MD) simulations are 
frequently used to model collision cascades during neutron 
bombardment in fission or fusion reactors \cite{Nor18}. 
Then, the resulting damaged samples are analyzed by 
Wigner-Seitz cell or voronoi diagram methods to quantify the 
number of Frenkel pairs (interstitials and vacancies) formed in
the cascade \cite{PhysRev.43.804,doi:10.1063/1.4849775,Atsuyuki,Stukowski2009,FIKAR201860}. 
Nevertheless, formation of complex defects and thermal
motion have not been well studied or modeled by these methods \cite{Sand_2013}. 
In addition, the better the inter-atomic potentials are, the better 
the MD simulation can predict the induced damage 
in crystalline materials. 

Machine learning potentials have been proposed to improve the 
accuracy on the modelling of point defects formation in damaged PFMs.
W. Szlachta et al. \cite{PhysRevB.90.104108} recently developed the 
interatomic potential based on the Gaussian
Approximation Potential (GAP) framework \cite{PhysRevLett.104.136403,PhysRevB.87.184115}
to investigate tungsten in the bcc crystal phase and its defects. 
However, this potential cannot be utilized in the study of material 
damage by neutron bombardment due to the lack of information of the 
repulsive region to treat realistic short distance interactions.

In the current work, we take into use a very recently developed machine learning 
W interatomic  potential based on GAP \cite{Jesper_GAP}, that includes physical 
properties relevant to collision cascades in the training data set.
Then, the descriptor vector (DV) based method is used to analyze 
the damaged material, which is capable to assign a probability of being 
a point defect to each atom in the sample \cite{Jav_UvT}.

Our paper is organized as follows: in Sec. \ref{methods} 
we discuss the development of a machine learned (ML) potential and
describe the DV based method, which is 
capable to identify, classify and quantify standard and 
uncommon material defects.
In Sec. \ref{results}, we present the analysis of point defects formation 
in W samples at the PKA range of $0.5-10$ keV. 
In order to provide an insight of 
the limitations and advantages of our new ML inter-atomic potential, we compare
our results to those obtained by commonly used EAM \cite{AcklandGJ}
and Tersoff-ZBL \cite{Juslin} potentials. 
Finally, in section \ref{conclusions}, we discuss our results and 
provide concluding remarks.

\section{Theory}
\label{methods}

Since tungsten has been used widely in fusion machines, our work 
is dedicated to study the formation of standard defects and the 
detection of unforeseen point defects in crystalline W material 
samples.

\subsection{Machine learned inter-atomic potential}
\label{sec:modified_GAP}

Machine learned (ML) inter-atomic potentials are
not restricted to an analytical form and can be systematically improved
towards the accuracy of the training data. 
In order to model collision cascades, the ML potential must be able 
to treat realistic short-range dynamics defined by its repulsive part. 
In addition, the correct structure of the liquid phase and 
re-crystallization process (accurate elastic energies) should be well 
described, to accurately emulate atomic mixing together with defect
creation and annihilation during the collision cascade.
In this work, we use a ML potential recently developed
\cite{Jesper_GAP} within the Gaussian Approximation Potential (GAP)
framework \cite{PhysRevLett.104.136403,PhysRevB.87.184115}. 
Here, the total energy of a system of $N$ atoms is expressed as 
\begin{equation}
    E_{\textrm{tot}} = \sum^N_{i<j} V_{\textrm{pair}}(r_{ij}) + 
    \sum^N_{i} E^i_{\textrm{GAP}},
\end{equation}
where $V_\mathrm{pair}$ is a purely repulsive screened Coulomb
potential, and $ E_{\textrm{GAP}}$ is the machine learning
contribution. $ E_{\textrm{GAP}}$ is constructed using a 
two-body and the many-body Smooth Overlap of Atomic Positions 
(SOAP) descriptor \cite{PhysRevLett.104.136403}, and given by

\begin{equation}
    \begin{split}
        E^i_{\textrm{GAP}} & = \delta_\textrm{2b}^2 \sum_j^{M_{\textrm{2b}}}
        \alpha_{j,\textrm{2b}} K_{\textrm{2b}} (\vec q_{i,\textrm{2b}}, 
        \vec q_{j,\textrm{2b}}) \\
         & + \delta_\textrm{mb}^2 \sum_j^{M_{\textrm{mb}}} \alpha_{j,\textrm{mb}}
         K_{\textrm{mb}} (\vec q_{i,\textrm{mb}}, \vec q_{j,\textrm{mb}}), \\
    \end{split}
\end{equation}
where $\delta$ is the standard deviation of the Gaussian process
that sets the energy ranges of the training data, which contains 
the energy information and is chosen by systematic convergence tests 
\cite{Jesper_GAP}; $K$ is the kernel function representing the similarity 
between the atomic environment of the $i$-th and $j$-th atoms;
$\alpha$ is a coefficient obtained from the fitting process; 
and $\vec q$ is the normalized descriptor vector of the local 
atomic environment of the $i$-th atom (See Sec 2.3).
In the computation of the ML potential the descriptors for two body, 2b, 
is utilized to take into account most of the inter-atomic bond energies, 
while the many-body, mb, contributions are treated by the SOAP descriptor.

The GAP method has been applied by Szlachta et al.
\cite{PhysRevB.90.104108} to develop a ML inter-atomic potential
for tungsten to reproduce the properties of screw dislocations and 
vacancies.
However, this potential lacks of information for the modeling of collision 
cascades (See Appendix A) as 
self-interstitial atoms formation, the liquid phase and a realistic 
repulsive interactions are not included in the authors' training data. 
The new ML potential developed in Ref. \cite{Jesper_GAP} included 
these types of structures, and is 
therefore suitable for performing MD simulations of modeling material damage 
due to neutron bombardment. 
The elastic response of bcc W is also included in the training 
data of the new ML potential, which is an important property 
to treat the re-crystallization of the highly affected target region
during a collision cascade. 
More details about the development of this new potential can be found in
Ref. \cite{Jesper_GAP}.
%

\subsection{MD simulations}
\label{sec:md_simulations}

In order to explore the advantages and limitations of our new ML inter-atomic 
potential, MD simulations are performed to emulate a neutron bombardment process
by using standard and well known inter-atomic potentials 
\cite{Bonny_2014}. Then, a comparison between the obtained results, under the 
same conditions, is carried out.
Therefore, we first define a simulation box as a pristine W lattice sample 
based on a body-centered-cube (bcc) unit cell with a lattice 
constant of $a = 3.16$ \AA{} \cite{SETYAWAN2015329}.
Then, the samples follow a process of energy optimization and 
thermalization to 300 K using a Langevin thermostat with a time
constant of 100 fs., due to the experiments of tungsten damaging
are done at room temperature
\cite{Wright_nucl_fusion,Herrmann_nucl_fusion}.
The MD simulation starts by assigning a chosen PKA 
energy in a range of 0.5-10 keV to a W atom, which is located at 
the center of the numerical box. 
We take into account ten velocity directions for each PKA: 
$\langle 0 0 1 \rangle$, $\langle 0 1 1 \rangle$, 
$\langle 1 1 1 \rangle$, and 7 cases for 
$\langle r_1 r_2 r_3 \rangle$ where $r_i$ are random numbers 
uniformly distributed in an interval of $[0,1]$.
In Tab. \ref{tab:tab1}, we present the size of the numerical boxes 
as the number of unit cells with a side length 
of $a$; the number of W atoms in each numerical box; and the time
step used in the simulations as a function of the PKA.
We utilize the Velocity-Verlet integration algorithm to model the 
collision dynamics, which is performed for 10 ps, 
followed by a relaxation process for 5 ps.
The MD simulations were done in a desktop computer by using the 
Large-scale Atomic/Molecular Massively Parallel Simulator (LAMMPS) 
\cite{PLIMPTON19951} with the Quantum mechanics and 
Interatomic Potential package (QUIP) \cite{quip} that is used 
as an interface to read machine learned inter-atomic potentials
based on GAP \cite{PhysRevLett.104.136403}. 
We also chose the following standard potentials:
The reactive inter-atomic potential for the ternary system W-C-H of 
Juslin et al. \cite{Juslin} referred as J-T-ZBL in our work, which is
based on an analytical bond-order scheme.
This potential has been used to study neutron damage in poly-crystalline 
tungsten \cite{1402-4896-2011-T145-014036}, trapping and dissociation processes of 
H in tungsten vacancies \cite{FU2018278}; and cumulative bombardment of low 
energetic H atom of W samples for several crystal orientations \cite{FU2018}.
In addition, we use the interatomic potential based on the embedded-atom method, 
EAM, or Finnis-Sinclair model with modification by Ackland et al. \cite{AcklandGJ} 
and repulsive fit done in Ref. \cite{zhong_defect_1998}, 
denoted as AT-EAM-FS in this work, that has been applied to study Frenkel pair 
formation as a function of the PKA in pristine tungsten \cite{SAND2016119} and 
self-sputtering of tungsten in a wide impact energy range \cite{SALONEN200260}.

\begin{table}[!t]
\centering
\begin{tabular}{c r r r}
\hline
\textbf{PKA} & \textbf{Num. atoms} & \textbf{Box size [$a$]} 
& $\Delta t$ (ps)\\
\hline
0.5 &  35  152   & (25, 25, 25) & $10^{-3}$ \\
1   &  35  152   & (25, 25, 25) & $10^{-3}$  \\
2   &  35  152   & (25, 25, 25) & $10^{-3}$ \\
5   & 124  722   & (38, 38, 40) & $10^{-4}$ \\ 
10  & 235  008   & (47, 47, 50) & $10^{-4}$ \\
\hline
\end{tabular}
\caption{Size of the numerical boxes based on a bcc unit cell 
as a function of the impact energy (PKA velocity), which is used
in the MD simulations. 
The box size is reported as the number of unit cells with side
length of $a=3.16$ \AA{}, that is the lattice constant of W at 
$300$ K.}
\label{tab:tab1}
\end{table}

\subsection{Descriptor vectors based method}
\label{sec:DV_method}

The quantification and classification of point defects in a damaged sample starts 
by computing the descriptor vector (DV) of all the atoms in the material sample.
The DV of the $i$-th atom of the sample, $ \vec{\xi}^{\ i}$ (defined below), 
is invariant to rotation, reflection, translation, and permutation of atoms
of the same species, but sensitive to small changes in the local atomic 
environment \cite{PhysRevB.87.184115}. 
It can be considered as a finger-print of the particular atomic environment of 
an $i$-th atom, which is expressed by a sum of 
truncated Gaussian density functions as  \cite{PhysRevB.87.184115},

\begin{equation}
\rho^{{} i}(\vec r) = \sum^{\textrm{neigh.}}_{j} \exp 
\left( -\frac{|\vec r-\vec r^{\ ij}|^2}{2 \sigma^2_{\textrm{atom}}} \right) 
f_{\textrm{cut}} \left( |\vec r^{\ ij}| \right),
\label{eq:Eq1}
\end{equation}
here $\vec r^{\ ij}$ is the difference vector between the atom
positions $i$ and $j$.
$\sigma^2_{\textrm{atom}}$ defines the broadening of the 
atomic position, which is set according to the lattice constant
of the sample.
Finally, $f_{\textrm{cut}} \left( |\vec r^{\ ij}| \right)$ is 
a smooth cutoff function, that limits the considered neighborhood of 
an atom.  
The function $\rho^{i}(\vec r)$ can also be defined in terms 
of expansion coefficients, $c_{nlm}$, that corresponds
to the $i$th-atom in the lattice  as \cite{PhysRevB.90.104108},
\begin{equation}
\rho^i(\vec r) = \sum_{nlm}^{NLM} c^{(i)}_{nlm}g_n(r)Y_{lm}\left(\hat r\right),
\label{eq:Eq2}
\end{equation}
where  $c^{(i)}_{nlm} = \langle g_n Y_{lm} | 
\rho^i \rangle$, $\hat r$ is a unit vector in the $\vec r$ direction,
$g_n(r)$ is a set of orthonormal radial basis functions $\langle
g_{n}(r)\mid g_{m}(r)\rangle = \delta_{nm}$, and $Y_{lm}(\hat r)$
are the spherical harmonics with the atom positions
projected onto a unit-sphere. 
Thus, Eq. \ref{eq:Eq2} is averaged over all possible rotations to 
be invariant against rotations, by the 
product of the $c_{n'lm}$ with its complex conjugate coefficient $c^{*}_{nlm}$, 
summed over all $m$.
Then the DV of the $i$-th atom, $ \vec{\xi}^{i}$, is defined as  \cite{PhysRevB.90.104108}
\begin{equation}
\vec{\xi}^{\ i} = \left\{ \sum_m
\left(c_{nlm}^i \right)^* c_{n'lm}^i \right\}_{\ n,n',l},
\label{eq:Eq3}
\end{equation}
where each component of the vector corresponds to one of the index triplets $\{n,n',l\}$.

In this work, we refer to the DV as the normalized vector $\vec{q}^{\ i} = 
\vec{\xi}^{\ i}/|\vec{\xi}^{\ i}|$ for the local environment of the $i$-th 
atom.
Depending on the choice of the expansion orders in Eq. \ref{eq:Eq2} 
for the spherical harmonics and the radial basis functions the 
number of components of $\vec{q}^{\ i}$ varies. 
In order to compute the DVs, we used the SOAP descriptor tool in QUIP with 
we consider a cutoff distance of $3.1$ \AA{}, which allows us to describe the 
local atomic environment and to identify lattice distortions and defects at the 
first nearest neighbors.
This parameter is chosen according to the W lattice constant at 0 K, $3.16$ \AA{}. 
The values for the spherical harmonicas are $N = 4$, and $L = 4$ (with $-L \leq m \leq L$) 
which yields to a vector with $k$ = 51 $\left( 0 \dots 50  \right)$ components. 

\subsection{Identification of point defects}
\label{sec:point_defects}

The difference of two local environments of the $i$-th atom and $j$-th atom 
can be computed by calculating the distance, $d$, between two DVs, 
$d = d \left( \vec{q}^{\ i}, \vec{q}^{\ j} \right)$.
However, we keep in mind that some vector components may be more
fluctuating than others and an appropriate measure to compare the 
DVs is done as follows: 
We define a small simulation box with hundreds of W atoms to 
apply a Langevin thermostat, which generates a thermalized tungsten
bcc lattice without defects to a desire sample temperature,
$T = 300$ K, in our case. 
Then, we compute the DVs of all the W atoms to calculate a mean 
reference DV,
$\vec{v}\left(T\right)=\frac{1}{N}\sum_{i=1}^{N}\vec{q}^{i}\left(T\right)$,
for defect-free environments; as well as the associated
covariance matrix, $\Sigma\left(T\right)$ , which highly depends on
the temperature of the sample. 
Therefore, the distance difference between a thermalized atomic 
environment and a damaged one is computed by the Mahalanobis 
measure as \cite{Maha}
\begin{equation}
    d^M (T) \left(\vec{q}^{\ i},\vec{v}\left(T\right)\mid \Sigma \right)=
    \sqrt{ \left( \vec{q}^{\ i} - \vec{v}\left(T\right) \right)^{\textrm{T}} 
\Sigma^{-1} (T) \left( \vec{q}^{\ i} - \vec{v}\left(T\right) \right)},
\label{eq:maha}
\end{equation} 
where $\left( \vec{q}^{\ i} - \vec{v}\left(T\right) \right)^{\textrm{T}}$ is 
the transpose vector.
This provides us information about the presence of an unexpected 
large distortion of the local environments. 
In order to detect common types of defects, a similar approach 
has been chosen at $T =0$ K.
A small simulation box containing the defect of interest
 (e.g. an interstitial, an atom next to a single vacancy) is
 prepared and the DVs of all the atoms are then computed, 
 subsequently acting as a fingerprint for this specific
 type of defect (see section 3).

The definition of the distance difference, $d^M$(T), between 
to local atomic environments conducts a probabilistic interpretation
of the obtained results. 
Thus, the probability, $P\left(\vec q^{\ i} \mid \vec{v}\left(T\right)\right)$, 
of an $i$-th atom being in a locally undistorted lattice can be computed using

\begin{equation}
\centering
P\left(\vec q^{\ i} \mid \vec{v}\left(T\right)\right)  = 
P_0 \exp \left[ -\frac{1}{2}d^M(T)^2 \right],
\label{eq:Eq4}
\end{equation}
where $P_0$ is the normalization factor.
Therefore, all the atoms in a damaged material sample have 
an assigned probability of being in a lattice position and 
atoms with the lowest probability will be labeled as point defects 
in the sample, following a type defect classification \cite{Jav_UvT}.
Here, atoms with lower probability define the distorted region around 
the permanent defects, which provides a good visualization of the 
damaged in the material

\section{Results}
\label{results}

In order to test the advantages and limitations of the 
new ML inter-atomic potential to traditional ones like J-T-ZBL and AT-EAM-FS 
potentials; we performed 
MD simulations at a primary knock-on atom (PKA) energy of 1 keV, 
in the $\langle 0 0 1 \rangle$ velocity direction, with a 
sample temperature of $300$ K.
Then, the damaged sample is analyzed by applying our DV based 
method to identify the formation of standard point defects (e.g. 
interstitial, vacancies) and unforeseen defects. 
The comparison of our results with those obtained by using 
AT-EAM-FS \cite{AcklandGJ} and J-T-ZBL
\cite{Juslin} potentials, under the same conditions, will serve as a 
test for our ML potential.

\begin{figure}[!b]
\centering
\includegraphics[trim={0 0 0 30pt},clip,width=0.5\textwidth]{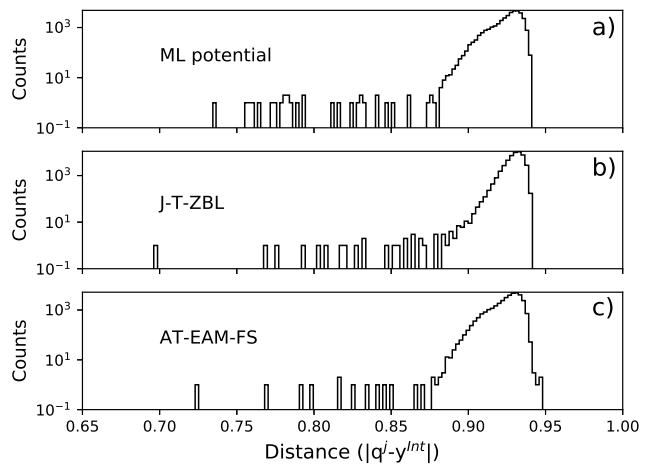}
\caption{Histogram of the distance difference between the 
interstitial DV and the W atoms in the damaged sample after 
relaxation process. 
MD simulation were performed by the ML potential in a), 
J-T-ZBL in our previous work \cite{Jav_UvT}, and 
the AT-EAM-FS potential in c). } 
\label{fig:fig0}
\end{figure}

In Fig. \ref{fig:fig0}, we present the distance difference 
between the reference DV vector of a W atom in an interstitial site to 
all the W atoms in the damaged sample after collision cascade. 
These results are obtained by Eq. \ref{eq:maha}, considering 
a reference DV of an interstitial site at $T = 0$ K, 
$\vec{v}_{\textrm{I}} \left(T=0\right)$.
We observe that the shape of the histograms 
in Fig. \ref{fig:fig0}a) and Fig. \ref{fig:fig0}b) are similar, 
however the results for the ML potentials present 
a clear distance gap between the W atoms in a 
lattice position and the ones in an interstitial site and the 
distorted region, at a distance difference of 0.88.
This makes the identification of the W atoms in 
the vicinity of the interstitial atoms simple, and serves as a good test 
for this new MD potential. 
The W atoms can recover to their lattice positions ( material re-crystallization) 
after collision cascade during the relaxation process due to the elastic energy
is well defined in the training data.
The histogram reported in Fig. \ref{fig:fig0}b) and in our 
previous work \cite{Jav_UvT} shows a narrow shape and the W atoms that are 
in the vicinity of the interstitial atoms are difficult to identify. 
Finally, our results are compared to those computed by the 
Wigner-Seitz cell analysis \cite{PhysRev.43.804}, which is 
implemented in OVITO \cite{Stukowski2009}.
Although, this analysis is limited by the definition of spatial 
region around the W atoms, we have a good agreement by finding 
the two Frenkel pairs (single vacancy and a single 
self-interstitial atom) formed at the location in the damaged W lattice. 
These W atoms, called SIA, have the lowest probability to be at a lattice 
position \cite{Jav_UvT}.
However, our method is capable to identify the W atoms that 
are in the vicinity of the SIAs, which is observed as a distorted region.
This visualization can be done \textit{via} OVITO and choosing different distance 
thresholds manually.

\begin{figure}[!t]
\centering
\includegraphics[width=0.5\textwidth]{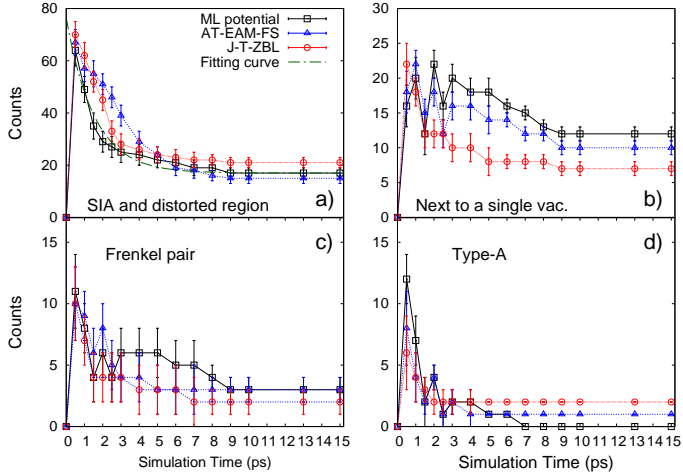}
\caption{Quantification and classification of material defects
formation as a function of the time for 10 MD simulations. 
We follow the formation of point defects during and after collision 
dynamics obtained by using the new ML potential, J-T-ZBL, and AT-EAM-FS. 
We add a fitting curve to the SIA and distorted region counting as 
$f(t) = f_0 \exp{\left( -\alpha t \right)}+\beta$ in a).}  
\label{fig:fig02}
\end{figure}

Since the formation of interstitials is well modeled by the 
new ML inter-atomic potential, according to our previous analysis.
It is interesting to investigate the formation of different point 
defects as a function of the simulation time by considering ten 
different velocity directions (10 MD simulations). 
In Fig. \ref{fig:fig02} we present the quantification and
classification of defects formation during collision dynamics (0-10 ps) by using 
the new ML potential (Fig. \ref{fig:fig02}a), 
J-T-ZBL (Fig. \ref{fig:fig02}b), and AT-EAM-FS (Fig. \ref{fig:fig02}c) 
in the MD simulations. 
The defects in the material remain during relaxation process (10-15 ps).
The total number of point defects after the collision cascade presented by 
the ML potential is similar to those performed with the standard potentials, 
under the same conditions. 
Nevertheless, W atoms tend to adapt to their interstitial site gradually 
during the collision cascade simulation. 
While the J-T-ZBL and AT-EAM-FS simulations show more W atoms as 
interstitial in the lapse time of 1-3 ps. 
In the same figure, we add a fitting curve to the number of SIA and 
atoms in the distorted region as: $f(t) = f_0 \exp{\left( -\alpha t \right)}+\beta$, 
with $f_0 = 31$, $\alpha = 0.65$ ps$-1$, and $\beta = 17.0$.
The second classification is defined as a W atom next to a single 
vacancy, $\vec{v}_{\textrm{V}}(T=0)$, in a bcc unit cell, at 0 K
\cite{Jav_UvT}. 
In Fig. \ref{fig:fig02} we notice that the number W atoms that belong to this 
classification is similar for the ML potential and AT-EAM-FS results.
A third classification is a type-A defect, $\vec{v}_{\textrm{A}}(T=0)$, 
which is defined as a W atom in the vicinity of a split vacancy or di-vacancy 
\cite{Jav_UvT}.
The formation of this defect is observed at the beginning of the MD simulations 
by the three inter-atomic potentials, however only the ML potential does not 
preserve this defect after collision cascade. 
It is known that this defect is energetically unstable according to DFT calculations
\cite{Heinola_2017}.
Therefore, it is important to notice that the formation of a type-A defect is less 
favorable for MD simulations that uses the machine learned potential \cite{Jesper_GAP}.
Providing an advantage of our MD potential over the standard ones.

In order to obtain a visualization of the spatial volume of the identified 
vacancies, we first define a sampling grid by 200 points of lateral dimension 
and a spatial step of 0.5 \AA{} that fits the numerical box of the damage sample. 
Then the nearest neighbor distance between the spatial position of the atoms and 
the grid points can be calculated by a k-d-tree algorithm
\cite{Bentley:1975:MBS:361002.361007,Jav_UvT} with the KDTREE2 code \cite{kennel2004kdtree}.
Then, squared distances larger than the lattice constant, 3.16 \AA {} for W, 
are used to identify the spatial volume around a vacancy in the damaged material.
Sampling grid points with the largest distances are associated to the location of 
a single vacancy.
In Fig. \ref{fig:fig02}c) we notice that at the beginning of the MD simulations 
the number of vacancies and W atoms next to single vacancy have the same trend. 
Then, the ML potential and AT-EAM-FS potential reach an agreement for the number 
of vacancies formed, however the J-T-ZBL results present a lower number of 
vacancies in the sample.
The detailed analysis of the material damage due to neutron bombardment by 
the DV based method and k-d-tree algorithm show that standard inter-atomic potentials 
have some unexpected errors during the modeling of collision cascades. 
In conclusion, the ML potential shows its first advantage at a PKA of 1 keV, 
regardless of the good agreement of the number of point defects with the standard 
potentials.

\subsection{Classification and quantification of crystal defects 
as a function of the PKA}

\begin{figure}[!b]
\centering
\includegraphics[width=0.5\textwidth]{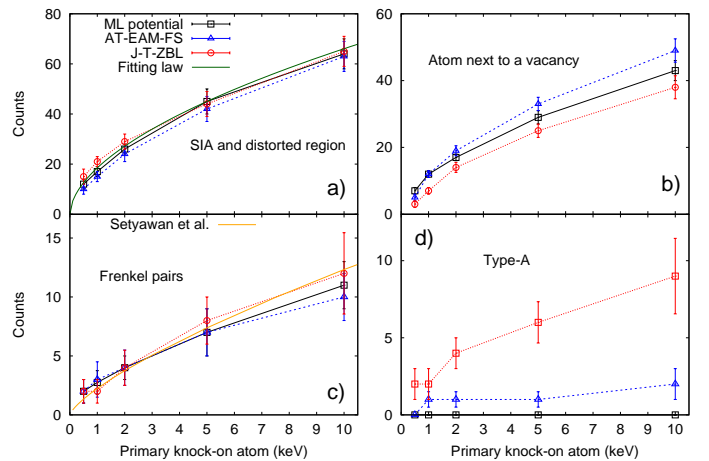}
\caption{Number of material defects as a function of the PKA energy, $E_{\textrm{PKA}}$. 
We include a fitting curve to the average number of SIA and atoms 
in its distorted region as:
Counts = $\alpha E_{\textrm{PKA}}^{\beta}$ \cite{BACON199537} with
$\alpha = 18.49$ and $\beta = 0.553$ with a  correlation factor of $0.99$. 
 Results at 1 keV were obtained in our previous work \cite{Jav_UvT}. 
 The number of vacancies are in good agreement with 
 the reported results by Setyawan et al. \cite{SETYAWAN2015329}  } 
\label{fig:fig3}
\end{figure}

\begin{table*}[!t]
\centering
\begin{tabular}{l rrrrr}
\hline \hline
\multicolumn{6}{c}{ML potential} \\
\hline 
& \multicolumn{5}{c}{PKA (keV)}\\
\hline
\textbf{Defect} & \textbf{0.5} & \textbf{1} & \textbf{2} & \textbf{5} 
& \textbf{10}  \\
\hline
Interstitial   & $12 \pm 2$ ($2 \pm 1$) & $17 \pm 2$ ($3 \pm 1$) & $26 \pm 3$ ($4 \pm 1$)  
& $45 \pm 5$ ($7 \pm 2$)  &  $64 \pm 6$ ($11 \pm 2$)    \\
Next to vac.   & $7 \pm 1$ & $12 \pm 1$ & $17 \pm 1$   & $29 \pm 2$   &  $43 \pm 3$     \\
type-A         & 0 & 0 &   0  & 0    &  0    \\
Total          & $19 \pm 2$ & $30 \pm 2$ & $43 \pm 3$ & $74 \pm 5$  & $107 \pm 7$     \\
\hline
Vacancy        & $2 \pm 1$ & $3 \pm 1$ & $4 \pm 1$    & $7 \pm 2$   & $11 \pm 2$     \\
\hline \hline
\multicolumn{6}{c}{J-T-ZBL} \\
\hline 
& \multicolumn{5}{c}{PKA (keV)}\\
\hline
\textbf{Defect} & \textbf{0.5} & \textbf{1} & \textbf{2} & \textbf{5} 
& \textbf{10 }  \\
\hline
Interstitial   & $15 \pm 2$ ($2 \pm 1$) & $21 \pm 2$ ($2 \pm 1$) & $29 \pm 3$ ($4 \pm 1$) 
& $44 \pm 5$ ($8 \pm 2$)  &   $65 \pm 6$ ($12 \pm 3$)   \\
Next to vac.   & $3 \pm 1$ & $7 \pm 1$  & $14 \pm 1$ & $25 \pm 2$  &   $38 \pm 3$   \\
type-A         & $2 \pm 1$ & $2 \pm 1$  & $4 \pm 1$  & $6 \pm 1$   &   $9 \pm 2$   \\
Total          & $20 \pm 2$ &  $30 \pm 2$ & $47 \pm 3$   &  $75 \pm 5$ &  $112 \pm 7$  \\
\hline
Vacancy        & $2 \pm 1$ & $2 \pm 1$ & $4 \pm 1$    &  $8 \pm 2$  & $12 \pm 3$   \\
\hline \hline
\multicolumn{6}{c}{AT-EAM-FS} \\
\hline 
 & \multicolumn{5}{c}{PKA (keV)}\\
 \hline
\textbf{Defect} & \textbf{0.5} & \textbf{1} & \textbf{2} & \textbf{5} 
& \textbf{10}  \\
\hline
Interstitial   & $10 \pm 2$ ($2 \pm 1$) & $15 \pm 2$ ($3 \pm 1$) &  $24 \pm 3$ ($4 \pm 1$)
& $42 \pm 5$ ($7 \pm 2$)  &   $63 \pm 6$ ($10 \pm 2$)    \\
Next to vac.   & $5 \pm 1$ & $10 \pm 1$ &  $17 \pm 2$  & $33 \pm 2$  & $49 \pm 4$     \\
type-A         &  0 & $1 \pm 1$ &   $1 \pm 1$  &  $1 \pm 1$  & $2 \pm 1$      \\
Total          & $15 \pm 2$ & $28 \pm 2$  &  $48 \pm 4$  &  $76 \pm 5$ & $114 \pm 7$   \\
\hline
Vacancy        & $2 \pm 1$ & $3 \pm 1$ & $4 \pm 1$ & $7 \pm 2$ & $10 \pm 2$    \\
\hline \hline
\end{tabular}
\caption{Average number of point defects and vacancies 
as a function of the PKA, which are identified by our DV based method. 
Interstitials are counted at the total number of SIA + atoms in its distorted region.
SIA are identified as W atoms with the highest probability to be in an interstitial 
site, reported into parentheses. 
Total number of defects is calculated as: Interstitials+Next to vac.+type-A
}
\label{tab:tab2}
\end{table*}

We calculate the number of material defects at different PKAs as a test 
of our ML potential and DV based method.
For this, we perform MD simulations for an impact energy range of 0.5-10 keV at 
10 different velocity directions to count the remaining material defects 
after 10 ps for of simulation time for collision cascade and 5 ps running for a
relaxation process. MD simulations at 10 keV of PKA are performed for 20 ps for 
collision cascade.
In Fig. \ref{fig:fig3}, we report the average of the number of 
Self-Interstitial-Atom (SIA) and atoms in its distorted region in a), W atoms next 
to a single vacancy in b), Frenkel pair formation in c), and type-A (W atom in the vicinity 
of a split-vacancy or di-vacancy \cite{Jav_UvT}) in d) defects as a function of the 
impact energy. 
Since, a fitting curve to the formation of material defects as atoms in 
SIAs in different metal materials has been proposed by Bacon et al. and Stoller 
et al. \cite{BACON199537,STOLLER200022}, where Counts = 
$\alpha E_{\textrm{PKA}}^{\beta}$; whith $E_{\textrm{PKA}}$ as the PKA energy, 
$\alpha$ and $\beta$ are fitting parameters. 
We apply the damped least-square method to our results for the number of atoms 
in a distorted region, obtaining the fitting parameters as $\alpha=18.49$ and 
$\beta=0.553$ with a correlation factor of $0.99$. 
Besides that, a Frenkel pair is a typical defect, where the formation of an 
SIA is related to the creation of a vacancy, thus we can compare to the fitting 
curve reported by Setyawan et al. \cite{SETYAWAN2015329}.
This fitting law is expressed as a function of a reduced 
cascade energy as: $0.49 \left(E_{\textrm{PKA}}/E_{\textrm{d}}\right)^{0.74}$
at a sample temperature of 300 K with $E_{\textrm{d}} = 128$ eV;
Although the authors performed MD simulations by using 
corrected semi-empirical potentials by Finnis and Sinclair 
\cite{AcklandGJ}, this fitting law is in good agreement with our results. 
In the table \ref{tab:tab2}, we report the average of the 
number of material defects as a function of the PKA for reference. 
Interstitials are counted as SIA and atoms in its local neighborhood, and 
only those atoms with the maximum probability are counted as interstitial
sites (Frenkel pair) reported into parentheses. 
In good agreement with the number of 
vacancies formation, quantified and identified by the k-d-tree method.
The total number of defects is defined as: Total = Interstitials+Next to vac.+type-A.

\begin{figure*}[!t]
\centering
\includegraphics[width=150pt,height=160pt]{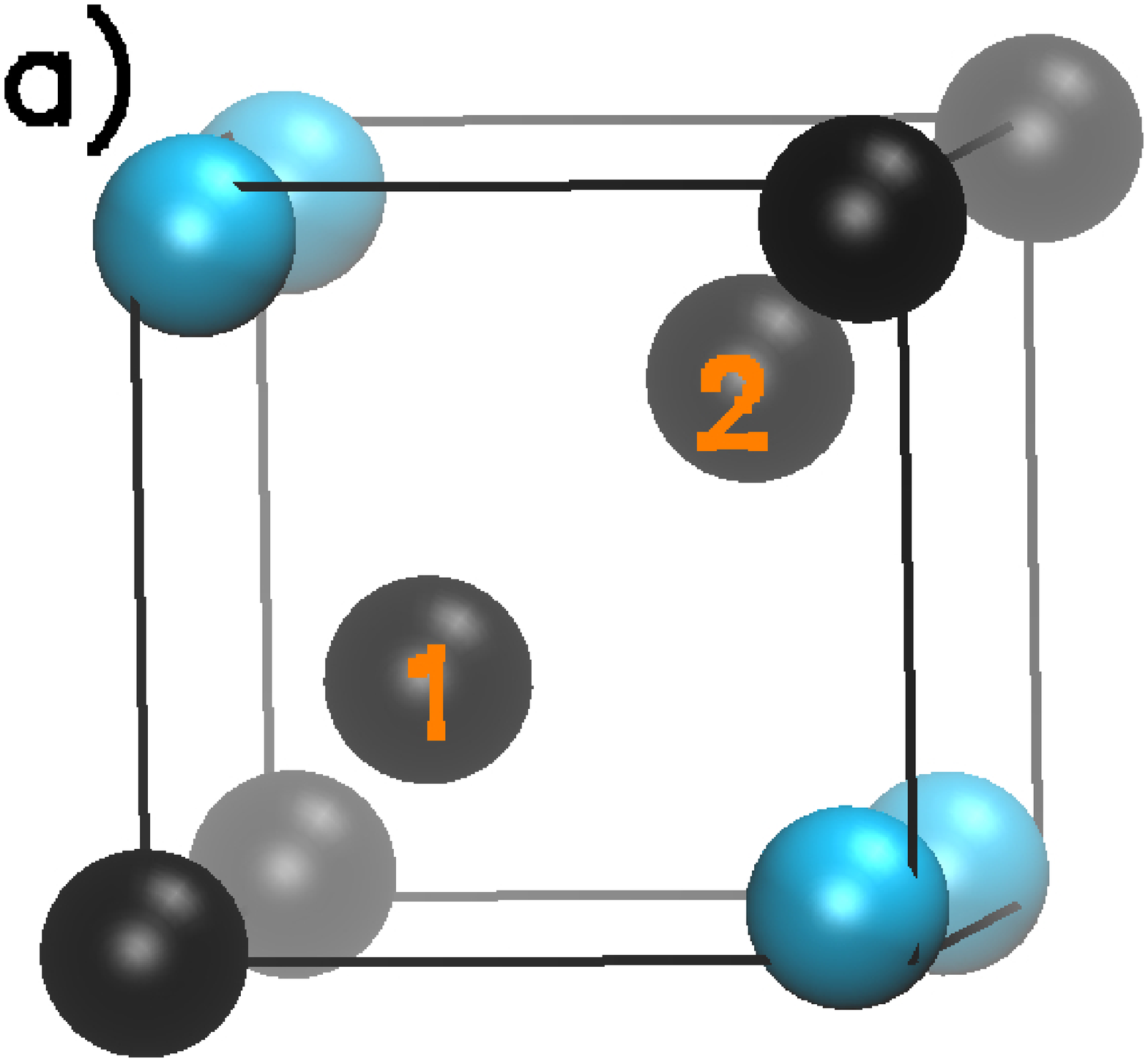}
\includegraphics[trim=0 0 0 50pt,width=220pt,height=150pt]{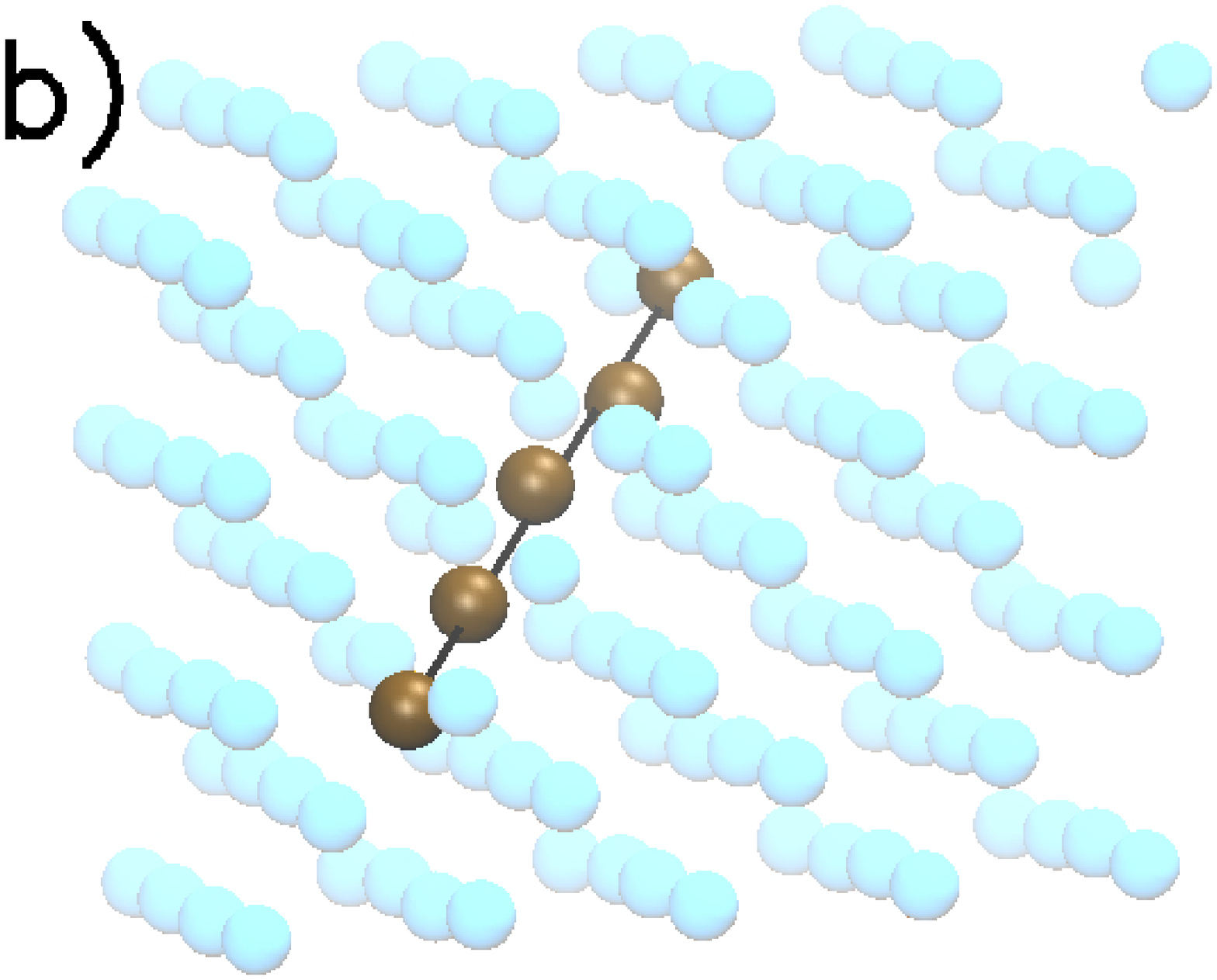}
\includegraphics[width=350pt,height=140pt]{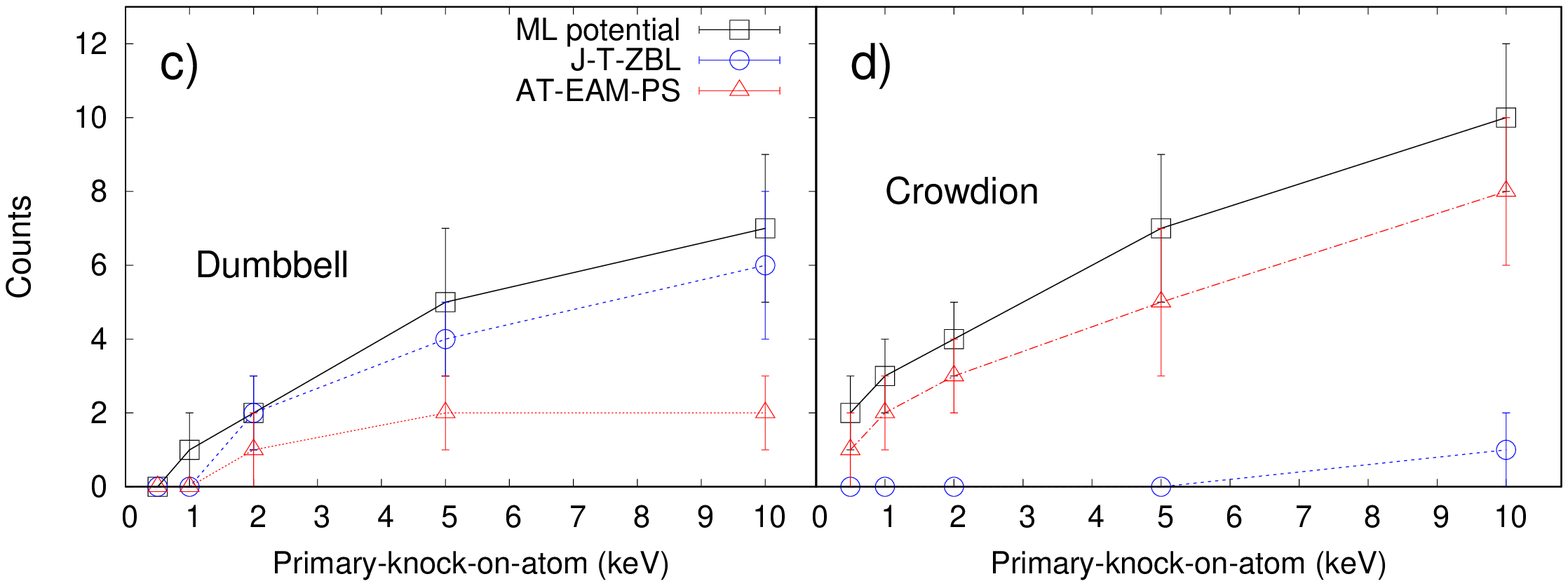}
\caption{A dumbbell defect is shown in a) and a crowdion line defect
is presented in b), identified at the final snapshot frame of the 
MD simulation with the ML potential at 2 and 1 keV of 
PKA, respectively. 
W atoms depicted as black (dumbbell) and golden (crowdion) spheres
represent the atomic arrangement of the defects and atoms in a 
lattice position are illustrated as blue (light-blue) spheres.
These defects are identified by the reference DV for an 
interstitial site, 
$\vec{v}_{\textrm{I}}(T=0)$ with a $\langle 1 1 1 \rangle$ 
orientation. 
The quantification of these defects is presented in c) and d). 
We compare results to those obtained by J-T-ZBL and AT-EAM-PS potentials.} 
\label{fig:fig4}
\end{figure*}

There is a couple of common and complex materials defects that are formed 
by several atoms in their interstitial sites.
A dumbbell defect, where two atoms share a lattice site, is the
most likely defect to be found in a bcc unit cell based materials
\cite{PhysRevB.73.020101,PhysRevMaterials.3.043606}, 
this type of defect oriented on $\langle 1 1 \xi \rangle$ with 
$\xi \approx 0.5$ is the most stable one according to DFT calculations 
\cite{PhysRevMaterials.3.043606}. 
It is well modelled by our new ML potential, and found in our MD 
simulations with an orientation of 
$\langle 1 1 \xi \rangle$ with $0.55 \leq \xi \leq 1$ due to the 
thermal motion.
In Fig. \ref{fig:fig4}a), we show the structure of a dumbbell 
defect found in our MD simulations after collision cascade at 2 keV of PKA; 
where W atoms represented by black sphere correspond to the dumbbell 
geometry and blue spheres are included to have a better visualization 
of this type of defect.
The average distance between the W atoms $1$ and $2$ is $2.18$ \AA{}.
This atomic arrangement is used to count the number of dumbbell
defects found in the W sample at different velocity direction 
and PKA.
Another common defect where four atoms share three lattice sites
is called a crowdion \cite{PhysRevB.76.054107}, which is stable
at the $\langle 1 1 1 \rangle$ direction and found it in 
our MD simulations at this orientation.
Fig. \ref{fig:fig4}-b) shows a snapshot of this particular 
material defect at the end of the MD simulation for a PKA of 
$1$ keV. 
W atoms illustrated as golden spheres represent the geometry of 
a crowdion defect, while W atoms depicted as light-blue spheres 
are considered as atoms in their lattice position. 
The average inter-nuclear distance between the W atoms that define a 
crowdion is $2.3$ \AA{}.
The geometries of the material defects are reported in the 
Supplementary material.
In Fig \ref{fig:fig4} c) and d), we report the number of dumbbell and 
crowdion as a function of the PKA obtained by using the ML  
potential \cite{Jesper_GAP}. 
A comparison to results given by MD simulations with the J-T-ZBL
potential shows the absence of crowdion defects formation at low 
impact energies, where the machine learned MD simulations predicts 
the formation of this type of defects in the whole PKA range. 
The higher the PKA value is, the bigger the number of crowdion 
defects is.
Besides that, the J-T-ZBL potentials are able to model dumbbell
defects, and its quantification agrees well with the results obtained 
by using the ML potential.
A second comparison to the results obtained by AT-EAM-FS for the
identification of these types of defects is presented in the same figure. 
The same number of crowdion defects, in average, is formed 
after collision cascade by the ML potential and AT-EAM-FS potentials.
Also, the formation of dumbbells defects is observed in the 
MD simulations by these two potentials, but the total number of 
defects is different. 


\section{Concluding Remarks}
\label{conclusions}

In this paper, we performed molecular dynamics simulations to 
emulate neutron bombardment on Tungsten samples in an 
impact energy range of 0.5-10 keV, and a temperature of 300 K.
For this, we use a new machine learning inter-atomic potential
based on the Gaussian Approximation Potential framework.
This new ML potential is accurately trained to the liquid phase, 
which is important to model the highly affected collision target; 
the short-range inter-atomic dynamics by including an accurate repulsive 
potential; and some samples to better model the re-crystallization 
of the molten region.
The damage in the W material sample is analyzed by the classification 
and identification of point defects with our descriptor vector based 
method, which is based on the calculation of the rotation and translation
invariant DV that describes the unique atomic neighborhood 
of each W atom in the sample.
Common point defects like self-interstitial-atoms and W atom next to 
a vacancy, and vacancy formation are quantified and classified as a function 
of the PKA energy. 
We found that the formation of W atoms as SIA and those in 
their distorted local environment follow a law of $18.49 E_{\textrm{PKA}}^{0.553}$ 
with $E_{\textrm{PKA}}$ is the PKA energy. 
 Point defects as dumbbell shapes and W atoms next 
 to a single vacancy are formed in the whole impact energy range. 
 Our results have, in average, a good agreement with reported 
 results by standard potentials. 
 However, some energetically unstable defects are corrected in the 
 training data for the ML potential to improve the accuracy of the MD 
 simulations. 
 Finally, these two methods are quite general and can be applied to 
 develop efficient machine learning inter-atomic potentials 
 for bcc metals and the damaged material samples are 
 analyzed by the DV based method, which is a future work 
 for our research group.

\section*{Acknowledgments}
F.J.D.G gratefully acknowledges funding from A. von Humboldt 
Foundation and C. F. von Siemens Foundation for research fellowship.
Simulations were performed using the LINUX cluster 
at the Max-Planck Institute for plasma physics

\section*{Appendix A. Test of original GAP potential }
\label{sec:appenA}

A machine learned inter-atomic potential for tungsten based on the 
Gaussian Approximation Potential formalism was developed by W. Szlachta 
et al. \cite{PhysRevB.90.104108}. However, it lacks of information about the
repulsive potential, so that the projectile is expected to travel freely when 
a primary knock-on-atom is assigned to it, in a MD simulation. 
The new ML potential \cite{Jesper_GAP} includes a realistic short-range repulsion 
to correctly simulate collision cascades. 
In order to test our new ML potential, we perform a MD simulation 
at 1 keV of PKA with a sample temperature of 300 K. 
The original GAP \cite{PhysRevB.90.104108}, Juslin et al. (J-T-ZBL) \cite{Juslin}, 
Ackland-Thetford (AT-EAM-FS) \cite{AcklandGJ}, and our ML potentials 
are used to compare the projectile trajectory as a function of the simulation time.

\begin{figure}[!t]
\centering
\includegraphics[width=0.5\textwidth]{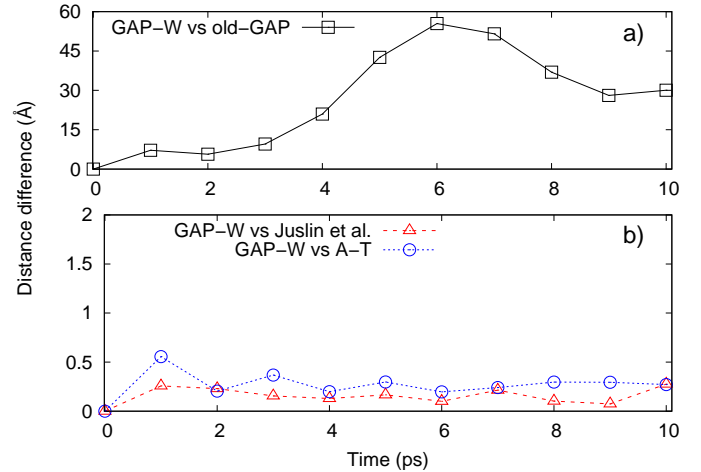}
\caption{Projectile trajectory comparison, as a function of the time, 
between the results obtained by new ML potential (GAP-W) and the one with 
the original GAP training data (old-GAP) in a), which shows the need of repulsion 
information in the training data set to model collision cascades.
We also compare the GAP results to those by Juslin et al \cite{Juslin} potential 
and Ackland-Thetford (A-T) potentials \cite{AcklandGJ} in b). } 
\label{fig:figA1}
\end{figure}

In Fig. \ref{fig:figA1}a), we present the comparison between the projectile trajectory 
calculated by the original GAP and the new ML potentials. 
The distance difference between two projectile trajectories is calculated as 
$\sqrt{\sum_i \left(\xi_i(t)-\Xi_i(t) \right)^2}$ where $\vec \xi(t)$ and 
$\vec \Xi(t)$ are the projectile trajectory obtained by different MD potentials, 
with $i =$ x, y, z.
We observe a notorious difference, where the projectile travels freely in 
the material sample during the whole simulation, as expected.
In Fig. \ref{fig:figA1}b), a similar comparison is done to the results obtained 
by using the J-T-ZBL and AT-EAM-FS potentials, as a function of the time. 
The distance difference is smaller than 1 \AA {} for the complete MD simulation 
and the final position of the projectile is the same for the three cases.
This result shows the well modelling of a collision cascade by our 
new ML potential.

\section*{Appendix B. Supplementary material}
\label{appendix:appendB}

\bibliography{bibliography}

\end{document}